# Role and Use of Race in AI/ML Models Related to Health

**Authors:**

Martin C. Were[1,2], Ang Li[3], Bradley A. Malin[1,4,5], Zhijun Yin[1,5], Joseph R. Coco[1], Benjamin X. Collins[1,6], Ellen Wright Clayton[6,7,8], Laurie L. Novak[1], Rachele Hendricks-Sturrup[9,10], Abiodun Oluyomi[3], Shilo Anders[1,5,11], and Chao Yan[1]

**Author Affiliations:**

[1]Department of Biomedical Informatics, Vanderbilt University Medical Center, Nashville, TN, United States
[2]Department of Medicine, Vanderbilt University Medical Center, Nashville, TN, United States
[3]Department of Medicine, Baylor College of Medicine, Houston, TX, United States
[4]Department of Biostatistics, Vanderbilt University Medical Center, Nashville, TN, United States
[5]Department of Computer Science, Vanderbilt University, Nashville, TN, United States
[6]Center for Biomedical Ethics and Society, Vanderbilt University Medical Center, Nashville, TN, United States
[7]Law School, Vanderbilt University, Nashville, TN, United States
[8]Department of Pediatrics, Vanderbilt University Medical Center, Nashville, TN, United States
[9]National Alliance against Disparities in Patient Health, Washington D.C., United States
[10]Margolis Center for Health Policy, Duke University, Durham, NC, United States
[11]Department of Anesthesiology, Vanderbilt University Medical Center, Nashville, TN, United States

**Corresponding Author:**

Martin C. Were, MD, MS
Professor
Department of Biomedical Informatics
Vanderbilt University Medical Center
Email: martin.c.were@vumc.org
Address: Suite 750, 2525 West End Ave, Nashville, TN, USA, 37203

**ABSTRACT**

The role and use of race within health-related artificial intelligence and machine learning (AI/ML) models has sparked increasing attention and controversy. Despite the complexity and breadth of related issues, a robust and holistic framework to guide stakeholders in their examination and resolution remains lacking. This perspective provides a broad-based, systematic, and cross-cutting landscape analysis of race-related challenges, structured around the AI/ML lifecycle and framed through "points to consider" to support inquiry and decision-making.

**INTRODUCTION**

The role and use of the social construct of race within health-related artificial intelligence and machine learning (AI/ML) models has become a subject of increased attention and controversy. As noted in the National Academies recent report "*Ending Unequal Treatment*", it is increasingly clear that race in all its complexity is a powerful predictor of unequal treatment and health care outcomes.[1] Appropriate inclusion of race within AI/ML models can identify differences in the outcomes of people with different backgrounds, creating opportunities for mitigation.[2] Yet, numerous examples exist of inappropriate inclusion of race or proxies of race in health-related models, which can harm large segments of the population.[3] Such effects have informed a growing number of recommendations to remove race from AI/ML models for health in several instances.[4-7] After describing racial and ethnic differences in health care, the NASEM committee recommended for the Department of Health and Human Services to support elimination of interventions that exacerbate health differences, and to ensure that tools and algorithms are equally valid and accurate for all people.[1]

The challenge, then, is on how to achieve this goal. In recent years, statistical and computations approaches and tools have been increasingly employed to identify and mitigate problems related to data representativeness and algorithmic fairness when it comes to use of race in AI/ML models.[8-10] Other bodies of work focus on characterizing what race represents within particular contexts, with an emphasis on optimizing health for all. These approaches also aim to elucidate how historical and existing social structures and practices affect health outcomes,[9,11] and advocate moving from race-based to race-conscious medicine.[12]

Developing and deploying AI/ML models that do justice to both computational and sociocultural aspects is challenging. Considerations of the quantitative and sociocultural factors related to race in AI/ML are complimentary. Quantitative factors typically emphasize numerical model accuracy and computational techniques to enforce similar model behavior across racial groups, whereas sociocultural considerations prioritize understanding of the root causes of undesirable differences, addressing ethical and societal norms and engaging with interested parties to consider the societal impact of models. Unfortunately, the current absence of a holistic framing of this topic makes it challenging for interested and affected parties to easily and systematically interrogate and address all relevant issues that surround role and use of race in AI/ML models related to health. In fact, individuals and teams with specific expertise risk approaching this subject from a narrow perspective that fails to consider the complexities, nuances, and potential trade-offs and conflicts involved.

Comprehensive and holistic guidance on the role of race and its use in AI/ML is needed. The primary goal of this paper is to identify, frame, and examine the broad range of issues that arise. This examination is conducted across the AI/ML lifecycle, identifying specific "points to consider" at each lifecycle stage. Issues cutting across the lifecycle are also highlighted. Framing the problem in this manner can enable key interested parties, such as racial group representatives, data collectors, developers, model auditors, model users, regulatory bodies, and policymakers to easily and comprehensively identify specific elements to examine and address for their particular use case, while being aware of the breadth of other related issues.

# AI/ML LIFECYCLE AS A FRAMEWORK TO EVALUATE ROLE AND USE OF RACE IN MODELS

The AI/ML lifecycle captures key steps involved in developing and implementing AI/ML models. Many variations of AI/ML lifecycles have been proposed.[13,14] While the steps incorporated in such lifecycles are similar, some variability exists.[15,16] In this paper, we rely on an AI/ML lifecycle centered around patients to frame the discussion around role and use of race in AI/ML models for health. This lifecycle has six steps, namely: (1) purpose; (2) population; (3) data; (4) model development; (5) model validation; and (6) model deployment (**Figure 1**). Steps in an AI/ML lifecycle are interdependent, with one step relying on earlier ones and informing those that follow. In general, earlier steps in the lifecycle influence the next step, but these connections are not necessarily unidirectional nor are they explicitly sequential. A later step in the lifecycle can affect what needs to be accomplished in earlier steps and vice versa – in **Figure 1**, this notion is represented by the narrower arrows flowing in the opposite direction.

The AI/ML lifecycle approach provides a framework to structure and analyze issues that arise when reasoning about the role and use of race and its application in AI/ML models at each step. Notably, several of the highlighted issues and considerations in this paper are not unique to the use of race in AI/ML. As such, a broad body of work is drawn upon to informing the topic at hand, underlining the value of various perspectives. This paper focuses on a breadth of considerations with relevance to the multiple interested parties.

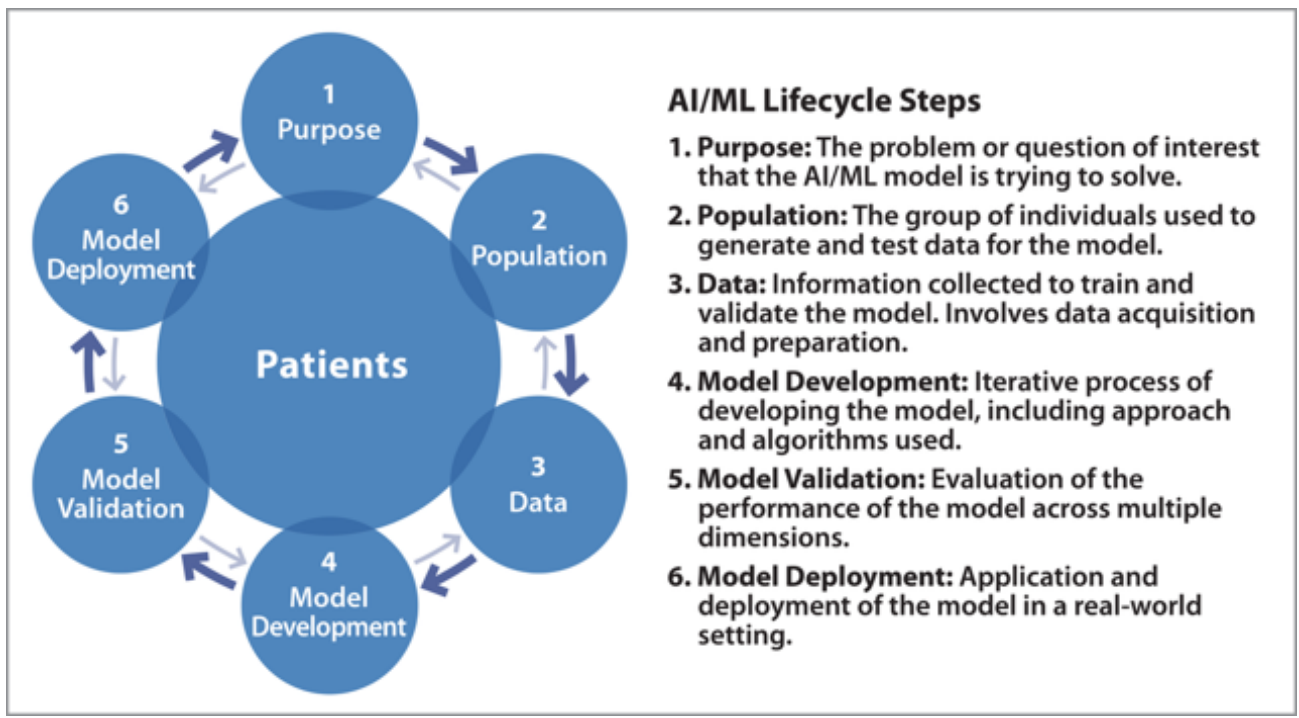

**Figure 1**: An AI/ML lifecycle model used to frame discussion on race, adapted from *Collins, et al.*[100].

# KEY CONSIDERATIONS FOR THE USE OF RACE IN THE AI/ML LIFECYCLE

**Purpose**

When it pertains to the use of race in AI models for health, the purpose of a model could be two-fold, namely: (a) a model that answers a non-race related question (e.g., develop a one-year mortality risk estimation model for all patients) but whose performance may differ across racial groups, or (b) a model that specifically evaluates a question or difference based on race (e.g., examine how cancer risk factors and outcomes differ by race). In both instances, the purpose that race serves in the model must be deliberately addressed. Race, being a social construct with no biological basis, must not be conflated with genetic differences, which often reflect ancestry.[17-20] It is now well-proven that race does not map to discrete genetic categories, and as such, differences observed by race in AI/ML models should not be assumed to arise from biological differences between races.[21]

AI/ML models should ideally meet the pressing needs of the target communities. In a world where some racial groups are more disadvantaged, under-resourced, and have multiple unmet healthcare needs, the question should be asked whether the purpose of the model meets the pressing needs of the affected racial groups. Yet approaches to systematically prioritize the needs of various groups are currently lacking. This area needs particular attention by policy- and decision-makers to ensure that AI/ML models respond to needs and optimize outcomes for all racial groups, and not just selected groups. It is also important to understand the relative risks and benefits of the AI/ML model for each racial group. While risk-benefit equation can and should be asked throughout the lifecycle, examining these early in the lifecycle can identify and mitigate issues before they arise and compound in effect. Where priorities between groups conflict or compete and where risks and benefits do not match among the groups, resolution via consensus-based approaches should be employed. **Table 1** highlights points to consider related to race and purpose of AI/ML models.

**Table 1:** Points to consider related to race and purpose of AI/ML models.

| Theme | Points to Consider |
|---|---|
| Genetic variation is not equal to race | • Do not blindly use race as a proxy for genetic variation in models. This requires being cognizant that models evaluating human genetic variation and ancestry do not use race as a proxy for genetic variation. |
| Interrogate what race represents | • Critically consider what race represents within a model, using findings to generate new hypotheses for examination as needed. |
| Prioritization of models | • Consider priority of the model being developed or implemented for all affected racial groups. |
| Consultative approach | • Gather inputs from relevant racial groups and systematically prioritize models for development and implementation that optimize benefits for all groups. |
| Address conflicts | • Address differences in risks and benefits as well as conflicts in interests between groups. |

**Population**

Population in **Figure 1** represents all categories of patients, research participants, community members, and other individuals from whom data are generated and used to train and test AI/ML models. Unfortunately, categorizing subsets of the population into racial groups can lead to misrepresentations and misconceptions when employed within AI/ML models. Two common misconceptions are that discrete race categories carry the same meaning across countries and that they remain unchanged over time. Yet definitions of racial categories can vary within and among countries.[22-24] Further, these definitions have historically changed over time, including the recent re-classifications by the Office of Management and Budget in the US that introduced a new race category of "Middle Eastern or North African", among other changes.[25,26] Individuals who do not self-identify with a single race also add complexity.[27,28]

Those from whom data are used in creating AI/ML models and on whom the models are implemented are not passive bystanders but rather are interested parties who directly experience the risks and benefits of developed models. Given the lack of public understanding of these, clear and proven community engagement strategies and collaborative partnerships that build trust must be employed before, during, and after implementation of AI/ML models.[29,30] For those who have less familiarity with these tools, this may require selecting appropriate community representatives to ensure that these groups have a voice and provide inputs into the process – akin to what is done is some consent scenarios.[31,32] As the target population may have important insights into what is at stake, these engagements can help to optimize mutual benefits and reduce disproportionate risks for particular racial groups throughout the model's development and deployment phases. Capacity-building initiatives will help these groups to better understand what is at stake as related to AI/ML models, support informed participation and sharing of data by these groups, and allow the groups to engage in highlighting areas where models do not apply accurately to them.[33]

Investigators from groups that have been less included in research can also provide valuable insights into the development and use of AI/ML. An example of such a capacity building and workforce development initiative is the 'Artificial Intelligence/Machine Learning Consortium to Advance Health Equity and Researcher Diversity (AIM-AHEAD)' that aims to increase participation and engagement of researchers and communities from all backgrounds in AI/ML initiatives.[29] **Table 2** highlights points to consider around race and populations on whom models are developed and implemented.

| Table 2: Points to consider around race and populations on whom AI/ML models are developed. | |
|---|---|
| **Theme** | **Points to Consider** |
| Meaning of racial categories | • Understand what various categories of race mean in the context of the model to be developed and whether these definitions have changed over time. |
| Generalizability of racial categories | • Examine generalizability of the racial categories used in developing the model, especially whether these categories apply similarly in different locations, countries, and time periods. |
| Engagement and collaborative partnership | • Employ appropriate community engagement and collaborative partnership strategies to inform all relevant stages of model development and to build trust. |
| Build capacity to comprehend AI/ML | • Build capacity among all racial groups to understand the role of AI/ML as well as specific relevant models and their implications. |

**Data**

The quality and quantity of the training data provided to a machine learning model has a major impact on its performance, such that inadequacies in the data can undermine the applicability of resulting models.[34] Incomplete or skewed collection of data from different populations can lead to flawed tools. The challenges of using non-representative data for racial groups have been broadly reported. An often-cited example is that of pulse oximetry devices that have been shown to perform worse for black patients than white patients – largely because these devices were trained on data from mostly white patients.[35-37] Even when various racial groups are represented in the data, the quality of their data, from the perspective of completeness, correctness, and freshness, often varies. As an example, in the US, data about race and ethnicity are more likely to be incorrect for non-white patients in administrative databases.[38,39] The proportion of missing racial data can also vary widely between racial groups within the same dataset.[39] In addition, the quality can be influenced by whether information is self-reported or recorded by observers (e.g., healthcare providers).[39-41] How data are labeled, including when automated approaches are used, can also introduce bias that adversely impacts certain racial groups.[42,43] Beyond issues originating from data themselves, inappropriate use of available data in model development (e.g., using medical costs as a proxy of a patient's health need for resources) can lead to consequences detrimental to certain subpopulations.[44]

Often, the differences observed between racial groups reflect other unaccounted factors such as social, economic, and environmental influences.[45,46] This notion is demonstrated in a model introduced by Segar and colleagues where prognostic performance for predicting in-hospital mortality for black patients improved when other non-medical drivers of health (NMDoH), such as location, income, wealth, language, and education, were added into the model.[47] Other studies have shown that adding NMDoH data to AI/ML models can help reduce errors in outputs and provide insights into some of the associated factors contributing to differences by race.[47-51] The question should therefore always be asked about whether NMDoH data can be used to augment or replace race in models.[52-54] In addition, incorporating genetic (ancestry) and other biological data when available can further improve models that might consider using race data.[18,27]

Given existing challenges around completeness and quality of race-based data within datasets, it is often necessary to ensure appropriate data collection and pre-processing approaches.[55] Beyond working towards the collection of more complete and representative data, statistical and computational approaches can be employed to recognize and, at times, mitigate data-related deficiencies. Common mitigation approaches related to data include: 1) removing race information from training data,[8,55-57] 2) adding relevant information as new variables,[58,59] 3) reweighting or rebalancing,[60] 4) removing disparate impact,[61] 5) learning fair representations,[62] and 6) developing or augmenting with synthetic data.[63] It should be recognized that simply discarding race from the equation can sometimes lead to greater harm.[64] A general guideline is to include race as a variable only when it can enhance model fairness and when there is a clear understanding of its role and meaning within the datasets. It is also important to note that no single approach will best improve fairness in all cases. Therefore, determining which data pre-processing approaches should be employed will depend on the particular AI/ML use case, ideally informed by individuals or teams with relevant expertise and by comprehensive evaluation obtained in subsequent stages of the lifecycle. **Table 3** outlines key pros and cons of each of these approaches.

| Table 3: Common data pre-processing approaches for mitigating racial bias in AI/ML models. | | | |
|---|---|---|---|
| **Approach** | **Description** | **Pros** | **Cons** |
| Remove race information[8,55-57] | Discard race as a variable from models to be developed. | Can prevent the perpetuation of race-based medicine that negatively impacts underserved subpopulations. | • Blindly and solely relying on this strategy (i.e., "fairness through unawareness") might negatively impact fairness when race correlates with unaccounted critical variations in health outcomes |
| Add relevant information as new variables[58,59] | Collect and incorporate important variables like NMDoH and relevant biological indicators or measures. | •Can oftentimes help to explain variations in patients' outcomes.<br>•Can mitigate or remove the independent impact of race in model outcomes. | • Might create redundancy, or induce noise if new variables carry invalid information. |
| Rebalance/reweigh existing data[60] | Randomly oversample underrepresented racial | •Balance representativeness and prevent majority | • No new information is introduced. |

| | groups or put more weight on these groups. | domination in model training.<br>• Low computational cost. | • Can cause overfitting and undermine generalizability. |
|---|---|---|---|
| Mitigate variable distinguishability[61] | Adjust the values of individual variables to make the relevant distributions across racial groups less distinguishable. | • Can effectively mitigate bias related to disparate impact. | • Can oversimplify complex relationships in the data.<br>• Might lose critical clinical information.<br>• Can reduce the overall accuracy.<br>• Might not generalize to other cohorts. |
| Learning fair representations[62] | Learn a latent representation for each data instance that obfuscates information about race. | • Can effectively mitigate differences in model performance related to disparate impact. | • Might lose critical clinical information.<br>• Can reduce the overall accuracy.<br>• Might not generalize to other cohorts.<br>• Can create difficulties for model troubleshooting. |
| Develop synthetic data[63] | Generate unseen data conditioned on protected attributes (e.g., race) and merge with real data for model training. | • Can enhance the representativeness of racial groups that are not well-represented in the data.<br>• Might improve fairness and overall model accuracy simultaneously. | • Synthetic data may not fully represent the complexity of specific use cases.<br>• Can amplify model performance differences in real data when inappropriately generated.<br>• Data creation can be resource intensive. |

With increased emphasis on explainable AI (XAI),[65-67] mechanisms should be set in place to highlight the provenance (origin and history) and lineage (path taken from original state to current state) of the race data used in AI/ML model.[68,69] This will help users to evaluate the quality and integrity of the data for the AI/ML model. Moreover, it can reveal whether the data were obtained ethically and comply with regulatory guidelines.

Use of dataset "nutrition labels," in particular, is increasingly being advocated. The dataset nutrition labels aim to establish standardized metadata that highlight the key ingredients of a dataset as well as unique or anomalous variables regarding distribution, missing data, and comparison to other "ground truth" datasets.[70] Labels related to race should detail the characteristics of different racial groups within a cohort. To support implementation of provenance and lineage of datasets, projects can leverage available metadata and data lineage tools.[68] **Table 4** summarizes key points to consider around data in informing use and role of race within AI/ML models for health.

| Table 4: Points to consider regarding race and the data used in AI/ML models. | |
|---|---|
| **Theme** | **Points to Consider** |
| Reliability of data source | • Determine the reliability of the data sources from which the racial data is derived. |
| Representativeness of data | • Assess whether data for all relevant racial categories are adequately represented to train the model and, if not, assess the feasibility of collecting more data for underrepresented subgroups. |
| Data labeling | • Evaluate the degree to which the race-based data were appropriately labeled. |
| Data pre-processing | • Apply appropriate approaches to handle data quality issues and to pre-process the data (**Table 3**). |
| Data provenance and lineage | • Gather and utilize provenance and lineage information on the data. |

**Model development**

In addition to the characteristics of the data underlying models, inappropriate outcomes of health-related AI/ML can also arise from the architectural design of the model.[54,71] To address both data and model challenges, a large number of approaches have been developed to enhance data and model quality during the model development stage.[8,71-74] These approaches acknowledge that algorithms are not impartial and that certain design choices by their architects can lead to better results in mitigating and addressing racial bias. Common types of algorithmic fairness include "individual fairness" (i.e., individual patients with similar data have similar likelihood of benefiting from the model), "counterfactual fairness" (i.e., the patient-level model outcomes are unaffected by variations in protected attributes such as race and other demographic information), and "group fairness" (i.e., model outcomes are similar across groups of sensitive attributes).[75]

Pertaining to race, group fairness is particularly relevant given its use in exploring the adequacy of application across demographic groups. Group fairness aims to define, quantify, and mitigate unfairness from AI/ML models that may cause disproportionate harm to certain subpopulations, such as to specific racial groups.[76] Numerous definitions of group fairness exist, each corresponding to a quantitative fairness metric that emphasizes a specific concern. Thus, the selection of fairness metrics should be based on the specific needs of each use case, recognizing that all metrics cannot be achieved at the same time.[77] Fairness metrics can be enforced during, as well as after, model training through the addition of non-discrimination constraints as part of the objective function.[71] While enforcing metrics can induce models that are more generalizable, the effectiveness of such approaches can vary and they could impact the overall model accuracy and introduce a higher level of complexity and cost for model implementation.[72,78,79] Moreover, enforcing fairness for one sensitive attribute (or one fairness metric) can inadvertently lead to unfair outcomes for another sensitive attribute (or another metric). As such, selection of the fairness enforcement strategy, including whether there is a need to do so, should be thoroughly assessed and tailored to specific use cases. A subset of available data needs to be set aside, using strategies like stratification and temporal selection, to conduct an initial evaluation of the model's accuracy and applicability across groups to provide feedback on the effectiveness of considered approaches for improving fairness. It should be noted, however, that directly applying these approaches can risk masking rather than resolving the deeper systemic issues that cause problematic applications, such as unequal access to healthcare or race-based patient treatment.

Given that race may correlate with social, environmental, and economic factors, appropriate approaches must be implemented during model development to handle such correlations when race is used as a covariate. At the very least, differences observed by race in AI/ML models should be scrutinized to better understand the exact cause(s) of the observed differences, which may involve other NMDoH. These observed differences should trigger hypotheses with subsequent examination to better understand the causes. Examination of variations within racial groups (within-group designs), using techniques such as hierarchical models, can provide insights into the causes of observed differences.[80,81] Further, when differences between racial groups are detected in models, a systematic approach should be applied to reduce differences between the groups in a unified model, while being attentive to not compromising performance.[82] However, if model performance is significantly affected in the unified model, it will be necessary to evaluate the implications of using different models by race or whether to consider other variables. Finally, attention should also be paid to whether models leverage embedded demographic information (such as race) as shortcuts to make predictions, even when race is not explicitly included as a variable.[83] Benefits of eliminating these demographic shortcuts and approaches to use will depend on the particular case. **Table 5** highlights points to consider during model development.

| Table 5: Points to consider regarding race during AI/ML model development. | |
|---|---|
| **Theme** | **Points to Consider** |
| Fairness Definition | • Determine the fairness definition(s) and corresponding metric(s) to pursue for the current use case. |
| Model selection and optimization | • Ensure that the selected model and optimization algorithm do not deliver outputs that some groups inappropriately. |
| Assess for fairness | • Before using any fairness enforcement approaches, determine if the trained models are unfair among racial groups (sub-group analysis) and identify the reasons for the observed unfairness. |

| Enforce fairness | - Compare and optimize fairness enforcement approaches in the model development stage. |
|---|---|
| Examine causes of differences | - Critically examine the various possible causes of difference by race in order to prevent inappropriate application of models. |
| Within-group analysis | - Perform within-group analyses. |
| Evaluate impact of fairness enforcement | - Assess the impact of fairness enforcement approaches on both fairness and model performance. |
| Unified versus distinct models | - When model performance for certain racial groups is unacceptably sacrificed for achieving fairness through a unified model, assess the ethical and technological feasibility of developing distinct models for different racial groups that can break out the tension between performance and fairness. |
| Embedded race information | - Determine if model uses embedded race information as shortcuts for factors such as NMDoH in decision making, and the implications of eliminating such shortcuts to best meet use case for the model. |

**Validation and Assessment**

Rigorous validation of model behavior should be conducted to ensure that the model performs as expected before deployment to ensure generalizability. This model validation and testing should be performed for both model performance and fairness across various scenarios, populations and under as many different constraints as possible. This is because the real-world environment in which the developed model will be deployed might differ from the data generation environment used during the model's development. While it is not uncommon for performance of a model to deteriorate from what was observed during development, recent findings have shown that the level of model performance achieved in a development dataset does not necessarily transfer to different datasets or application settings.[83] Examples of such discrepancies include variations or inconsistencies in 1) the demographics, NMDoH, and clinical characteristics of patient cohorts, 2) the availability of variables, 3) measurement techniques like medical devices and their algorithms, 4) clinical care protocols, and 5) data collection and labeling procedures.

Models developed in one region or country might not translate to another without proper modifications. Considering all these complexities, implementing a silent-mode pre-deployment validation, which mimics site-specific settings without showing results to end-users,[14] could be the optimal strategy for ensuring the robustness and effectiveness of the model before it goes live.[84] Ideally, additional measures beyond performance and algorithmic fairness, such as the impacts on care quality, eligibility, cost, and outcomes, should be thoroughly assessed across the various racial groups as part of pre-deployment assessment.[85,86] The cost-benefit ratio of different AI/ML interventions becomes particularly relevant given the close connection of race with differences in health-related outcomes across racial groups. In particular, the cost-benefit of an AI/ML model should be compared against other models, as well as against other proven interventions and approaches to inform which model should be considered for use relative to alternative interventions. Model assessment should also incorporate the feasibility of adoption, given the multiple infrastructure, financial, and human-resource constraints faced by various populations and settings. It might not be justifiable to advocate for deploying models that are too costly to deploy to groups with limited resources without requisite measures to assure success in implementation and outcomes. **Table 6** summarizes key considerations surrounding validation and assessment of models.

| Table 6: Points to consider regarding pre-deployment assessment of AI/ML models. | |
|---|---|
| **Theme** | **Points to Consider** |
| Pre-implementation validation | • Conduct rigorous validations on model performance and fairness before deployment. |
| Outcomes and risk assessments | • Assess whether the impacts of the model on outcomes and risk allocation are acceptable. |
| Feasibility assessment | • Conduct feasibility assessments on implementation success by sorting out the disparities associated with race. |
| Cost-benefit evaluation | • Examine cost-benefit analysis results of the model. |
| Comparative cost-benefit | • Compare the cost-benefit of the model against other proven interventions. |

**Model Deployment**

All implemented AI/ML models should be audited prior to deployment and monitored once deployed.[87] Even when a model does not have a race variable, it can still generate unfair outcomes because of potential correlations between race and other variables. Efforts to improve explainability of AI (XAI) can support decision-making on which AI/ML models an organization should deploy.[88,89] Of particular relevance are external audits of algorithms, which often require deploying organizations to work closely with model developers.[90,91] Continuous monitoring of deployed models is essential given that data and model drift can have significant impact on model performance and fairness across groups. By employing processes and methods to detect drift, organizations can identify models that need updating or discontinuation.[92] Like other informatics-based interventions, AI/ML models can have unintended consequences, which must be monitored and mitigated using various available approaches.[93-95] Unintended consequences can further be ameliorated through awareness of the interactions between model outputs and the users of the model. This will reduce model outputs from being incorrectly interpreted by the users who often have their outlook.

Deliberate application of principles to assure optimal outcomes for all can further uncover and mitigate negative impacts of AI/ML models that incorporate race. Well-accepted approaches, such as those by Whitehead and Dahlgreen,[96] are particularly applicable and can be adopted for AI/ML models being deployed. These would include a requirement for AI/ML models to: "(a) level up, and not level down; (b) improve the status of those who are disadvantaged; (c) narrow the health divide; (d) reduce social inequities throughout the whole population; (e) tackle the fundamental social determinants of health; and (h) facilitate equal access to services and ensure that particular racial groups do not pay more to access the tools than others".[96] As appropriate, distributive justice approaches that emphasize allowing all people to achieve their optimal health and resource allocation across the various racial groups should also be employed.[51] **Table 7** summarizes key considerations in deploying models when race is considered.

| Table 7: Points to consider regarding race and deployment of AI/ML models. | |
|---|---|
| **Theme** | **Points to Consider** |
| Deployment context | • Ensure context within which the model is being deployed is appropriate for that model. |
| Site-specific model assessment | • Evaluate performance of the model for various groups within the specific deployment setting. |
| External model audit | • Models need to be independently audited prior to deployment. |
| Monitor data and model drift | • Implemented models should be monitored to detect performance changes, and to inform updates needed or need for model discontinuation. |
| User awareness | • Maintain vigilance on how users interact with models and interpret the model's outputs. |
| Unintended consequences | • Monitor and mitigate unintended consequences. |
| Outcomes for all | • Use accepted frameworks to evaluate impacts of the AI/ML model on optimal access to health care for all. |

**Cross-cutting Considerations**

In addition to issues arising at each stage of the lifecycle, there are several cross-cutting issues regarding the role and use of race across the AI/ML lifecycle that deserve particular attention.

*Teams*: Teams with different types of expertise are involved at the various stages of the AI/ML lifecycle. As pertains to models that involve patients with multiple races, individuals with various backgrounds in teams can bring different and relevant insights and perspectives at each stage. Beyond community engagement and engagement with community representatives, deliberate capacity building and involvement of individuals with diverse backgrounds is also relevant for developers and implementer teams of these models. Teams also need to bridge computational and social-cultural aspects of model development and implementation by incorporating multi-disciplinary team members.

*Governance*: Governance mechanisms that ensure that data are obtained and used ethically, and approaches for the adoption and monitoring of race-based AI/ML models must be in place. Unlike medicines and devices that are often tightly regulated, regulation of AI/ML models is nascent at best,[97] but the pervasiveness of race-biased predictive models in broad use calls for extra vigilance when AI/ML models can variably impact the various racial groups require robust governance.[44,98]

*Organizational capabilities:* Institutions that serve disadvantaged groups are less likely to have the organizational capabilities to develop, implement, and monitor AI/ML models and applications.[99] Costs across the AI/ML lifecycle are often prohibitive, which can impede development and use when requisite human, financial, and infrastructure resources. Understanding and narrowing resource and capability gaps across institutions will help ensure that AI/ML benefits are derived by all groups.

*Evaluation*: To assure high quality models, evaluation must be incorporated at every step in the lifecycle. Evaluations across the lifecycle can range from adequacy of community engagement strategies, quality assessment of data, evaluations performance of the model, model generalizability, impacts on health outcomes, ethical considerations, cost-benefit, and acceptability to those affected, among others. These evaluations can uncover gaps and inform mitigation strategies.

**CONCLUSION**

The role and use of race in AI/ML models for health will continue to elicit debate and is one deserving further research and examination. At the very least, caution must be exercised when considering issues surrounding role and use of race within AI/ML models or in interpreting differences in model outputs based on race. This work provides broad-based guidance to those wrestling with this topic at any of the stages of the AI/ML lifecycle and should stimulate renewed and comprehensive scrutiny on role and uses of race within AI/ML models for health.

**ACKNOWLEDGEMENTS**

This work was led by the Applied Ethics AI sub-core within the AIM-AHEAD program's Infrastructure core. The research reported in this paper was supported by AIM-AHEAD Coordinating Center, award number OTA-21-017, and was, in part, funded by the National Institutes of Health Agreement No. 1OT2OD032581.

**AUTHOR CONTRIBUTIONS**

M.C.W. and C.Y. conceived the conceptual framework, conducted the literature search, and drafted the initial manuscript. A.L., B.A.M., Z.Y., J.C., B.X.C., E.W.C., L.N., R.H., A.O., and S.A. contributed expert knowledge and participated in editing, revising, and reviewing the manuscript. All authors approved the final version.

## COMPETING INTERESTS

All authors report no competing interests to declare.